\documentclass[prl,aps,floats,letterpaper,floatfix,footinbib,superscriptaddress,preprintnumbers,twocolumn]{revtex4}
\usepackage{amsmath}
\usepackage{color}
\usepackage[mathenv]{}
\usepackage{graphicx}
\usepackage{ulem}
\usepackage{natbib}
\begin{document}


\title{Generation of a stable low-frequency squeezed vacuum field with periodically-poled KTiOPO$_4$ at 1064 nm}

\author{Keisuke Goda}
\affiliation{LIGO Laboratory, Massachusetts Institute of Technology, Cambridge, Massachusetts 02139}

\author{Eugeniy E. Mikhailov}
\affiliation{The College of William $\&$ Mary, Williamsburg, Virginia, 23187}

\author{Osamu Miyakawa}
\affiliation{LIGO Laboratory, California Institute of Technology, Pasadena, California 91125}

\author{Shailendhar Saraf}
\affiliation{Rochester Institute of Technology, Rochester, New York 14623}

\author{Stephen Vass}
\author{Alan Weinstein}
\affiliation{LIGO Laboratory, California Institute of Technology, Pasadena, California 91125}

\author{Nergis Mavalvala}
\affiliation{LIGO Laboratory, Massachusetts Institute of Technology, Cambridge, Massachusetts 02139}

\begin{abstract}
We report on the generation of a stable continuous-wave
low-frequency squeezed vacuum field with a squeezing level of
$3.8\pm0.1$ dB at 1064 nm, the wavelength at which laser
interferometers for gravitational wave (GW) detection operate, using
periodically poled KTiOPO$_4$ (PPKTP) in a sub-threshold optical
parametric oscillator. PPKTP has the advantages of higher
nonlinearity, smaller intra-crystal and pump-induced seed
absorption, user-specified parametric down-conversion temperature,
wider temperature tuning range, and lower susceptibility to thermal
lensing over alternative nonlinear materials such as MgO doped or
periodically poled LiNbO$_3$, and is, therefore, an excellent
material for generation of squeezed vacuum fields for application to
laser interferometers for GW detection.
\end{abstract}

\maketitle

Injection of squeezed vacuum states into the output port of
laser interferometers for gravitational wave (GW) detection is a promising technique for
enhancing the sensitivity of future detectors, such as Advanced
LIGO \cite{adligo}, which are expected to be operational in the next
few years. Large squeeze factors in the GW band (10 Hz to 10 kHz),
and stable, long-term operation are among the key requirements that a
squeeze source must satisfy in order to be used in long baseline GW
detectors with large duty
factor \cite{abramovici1992science}. Squeezed
vacuum fields are typically generated using nonlinear crystals in
sub-threshold optical parametric oscillators (OPOs). Consequently,
the choice of nonlinear material is critical, since it determines
several important parameters, such as nonlinearity, phase-matching
type, absorption loss, pump-induced-seed absorption loss, laser
damage threshold, photo-refractive damage threshold, and
susceptibility to thermal lensing. Moreover, since all GW detectors
presently use high power Nd:YAG lasers sources at 1064 nm,
generating squeezed vacuum states at 1064 nm is also essential.

Presently, squeezed state sources for GW detectors at 1064 nm use
OPOs comprising magnesium oxide (MgO) doped lithium niobate
(LiNbO$_3$) crystals~\cite{mckenzie2004prl,vahlbruch2006prl}.
LiNbO$_3$ is a widely used nonlinear material with good long-term
stability. MgO doping increases its laser damage threshold and
reduces the effect of green-induced infrared absorption (GRIIRA)
\cite{furukawa2001applphyslett}, but also increases impurity and
inhomogeneity in the crystal, thereby increasing intra-crystal
absorption and scattering losses, which place a limit on the
attainable level of squeezing. In this Letter, we report on the
generation of a squeezed vacuum field using an alternative material,
periodically poled KTiOPO$_4$ (PPKTP) at 1064 nm.

KTiOPO$_4$ (KTP) offers several advantages over LiNbO$_3$. It has a
higher laser damage threshold, higher resistance to photo-refractive
damage, and lower susceptibility to thermal lensing. These
properties are especially critical for the long-term stability of
OPOs pumped by cw visible
sources~\cite{fan1987applopt}. KTP has an effective
nonlinear coefficient, $d_{\rm eff}$, comparable to LiNbO$_3$ at
1064 nm, and has been shown to generate relatively high levels of
quantum correlations \cite{laurat2003prl, feng2004optlett}.
Furthermore, recent progress in the electric field poling of
flux-grown KTP has made periodically poled KTP (PPKTP) an even more
promising candidate. With its high nonlinearity of $|d_{\rm
eff}|\simeq 10.8$ pm/V, PPKTP is a competitive alternative to
periodically poled LiNbO$_3$ (PPLN) and LiTaO$_3$ (PPLT).
Poling-induced losses of PPKTP are much lower than those of PPLN. For these reasons, higher
squeezing levels are attainable with PPKTP at realistically
available pump powers.

Since birefringent phase matching is not used in periodically poled
crystals, temperature tuning of their refractive indices in the
ordinary and extraordinary axes is not required. The only
requirement on the temperature control is to maintain the grating
period against thermal expansion. The typical FWHM of the effective
parametric down-conversion temperature for PPKTP at 1064 nm is
$5^\circ$C, whereas it ranges from a few tens to several hundred mK for
LiNbO$_3$ (depending on its length). This property is particularly
vital for the long-term stability requirement of GW observatories,
since the system is robust against temperature variations, unlike
the birefringent phase-matching case. The low phase-matching temperature (typically around room temperature) and large effective
temperature range of PPKTP also significantly reduce difficulties
in fabrication of ovens for temperature control.

Observation of squeezing in the GW band has been reported by
McKenzie et al \cite{mckenzie2004prl} and Vahlbruch et al
\cite{vahlbruch2006prl}. Both groups used MgO:LiNbO$_3$, achieving
squeezing levels of about 3 to 4 dB. Here we demonstrate the generation
of a squeezed vacuum field at 1064 nm using PPKTP at 1064 nm. This
is, to our knowledge, the first time that squeezing at 1064 nm using
PPKTP is reported. This squeeze source has all the advantages of
stability and ease of operation that PPKTP offers, at a wavelength
suitable for GW detectors.

PPKTP has previously been used in a cw OPO to generate squeezed
states, but not at 1064 nm. Aoki et al. reported the generation of
squeezed light in sideband modes of cw light at 946 nm using a PPKTP
crystal in an OPO and observed the squeezing level of $5.6 \pm 0.1$
dB and the anti-squeezing level of $12.7 \pm 0.1$ dB
\cite{aoki2005optexp}. More recently, Takeno et al. observed $9.01
\pm 0.14$ dB of cw squeezing at 860 nm using a PPKTP crystal in a
bow-tie cavity \cite{takeno2007arxiv}. Hirano et al. also observed
the generation of pulsed squeezed light from a single-pass
degenerate optical parametric amplifier (OPA) with a PPKTP crystal
pumped by a cw second-harmonic field and reported the squeezing
level of 3.2 dB and the anti-squeezing level of 6.0 dB at 1064 nm
\cite{hirano2005optlett}. To the best of our knowledge, these are
also the highest levels of pulsed and cw squeezing ever obtained.

\begin{figure}[t]
\includegraphics[angle=0, width=1.0\columnwidth]{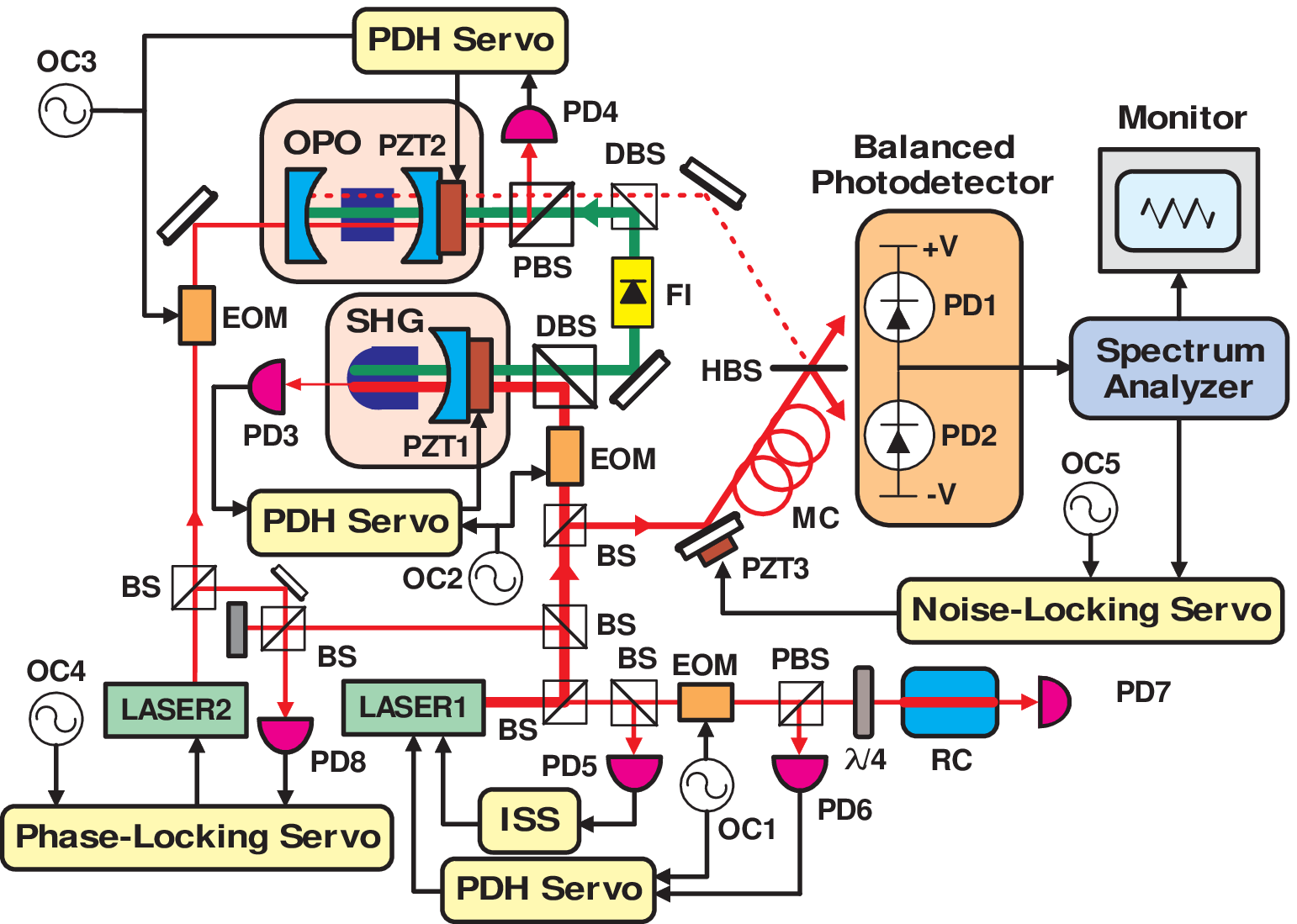}
\caption{(Color online) Schematic of the experiment. LASER1: Nd:YAG
MOPA laser; LASER2: Lightwave 126 laser; SHG: second-harmonic
generator; OPO: optical parametric oscillator; FI: Faraday isolator;
MC: mode-cleaning fiber; EOM: electro-optic modulator; PZT1, PZT2,
and PZT3: piezo-electric transducers; DBS: dichroic beamsplitter;
PBS: polarizing beamsplitter; HBS: 50/50 homodyne beamsplitter; BS:
beamsplitter/beam sampler; RC: reference cavity; $\lambda/4$:
quarter-wave plate; PD1, PD2, PD3, PD4, PD5, PD6, and PD7:
photodiodes; OC1, OC2, OC3, OC4, and OC5: local oscillators (13.3
MHz, 16.7 MHz, 49.5 MHz, 624 MHz, and 13.2 kHz, respectively); ISS:
intensity stabilization servo. Solid lines: coherent fields; dotted
line: squeezed vacuum field. A spectrum analyzer (Stanford Research
Systems SR785, not shown) is used to measure the balanced PD
signals.} \label{apparatus}
\end{figure}

The schematic of the experiment is shown in Fig. \ref{apparatus}.
The apparatus is mainly composed of (i) the second-harmonic
generator (SHG), (ii) the optical parametric oscillator (OPO), (iii)
the homodyne detector, (iv) the frequency- and intensity-stabilizer,
and (v) the phase-locked subcarrier (LASER2). The main light source is a
Nd:YAG Master Oscillator Power Amplifier laser (LASER1). The laser
is frequency- and intensity-stabilized using the LIGO pre-stabilized
laser system \cite{rollins2004optlett}.

The SHG is a cavity composed of a 6.5 mm long $5\%$ MgO:LiNbO$_3$ crystal
(Photon LaserOptik Inc.) with an anti-reflection coated flat surface
and a high-reflective curved surface and an output coupling mirror
with a radius-of-curvature of 50 mm. The radius-of-curvature and
reflectivity of the curved surface of the crystal are respectively 8
mm and 99.95$\%$ at both 1064 nm and 532 nm. The output coupler,
with reflectivities of 95.0$\%$ at 1064 nm and 4.0$\%$ at 532 nm,
is mounted on a piezo-electric transducer (PZT) as an actuator that
controls the cavity length. The SHG cavity is locked by the
Pound-Drever-Hall technique. The SHG cavity is enclosed in an oven
and the crystal is maintained at 114.05 $^{\circ}$C to optimize the
SHG conversion efficiency. The SHG cavity is pumped by 1.1 W of the
1064 nm from the MOPA laser and generates 320 mW at 532 nm, which is
used as a pump field in the OPO cavity.

The OPO is a cavity composed of a 10 mm long PPKTP crystal (Raicol
Inc.) with anti-reflection coated flat surfaces and two coupling
mirrors with a radius-of-curvature of 10 mm. The reflectivities of
the input and output coupling mirrors are 99.95$\%$ at both 1064 nm
and 532 nm and 92.0$\%$ at 1064 nm and 4.0$\%$ at 532 nm
respectively. The OPO cavity length is 2.3 cm. The crystal is
maintained at 33.5 $^{\circ}$C to optimize the 1064/532 parametric
down-conversion. When pumped by the second-harmonic pump at
frequency $2\omega_0$ and operated below threshold, the OPO
correlates the upper and lower quantum sidebands of a vacuum field
that enters the OPO around the center frequency $\omega_0$
\cite{caves1985pra}. The correlation of the quantum sidebands
appears as a squeezed vacuum field. A second laser beam (LASER2)
that is shifted by 642 MHz with respect to $\omega_0$ is used to
lock the OPO cavity in a TEM$_{00}$ mode. This frequency-shifted
light is orthogonally polarized to the vacuum field that seeds the
OPO cavity in a TEM$_{00}$ mode, and to the pump field.

The generated squeezed vacuum field is triggered by the coherent
local oscillator (LO) field and measured with a homodyne efficiency
of 99$\%$, using a balanced detector comprising a 50/50 beamsplitter
(HBS) and a home-made balanced photodetector that has two ETX500T
photodiodes (JDS Uniphase) with matched quantum efficiencies of
93$\%$. The difference between the two optical responses is
amplified and measured by the spectrum analyzer (HP8591E). The HBS
and balanced PD form an opto-electrical Mach-Zehnder interferometer
to extract the linear response of the squeezed vacuum field. The
squeeze angle is locked by the noise-locking technique
\cite{mckenzie2005job}, and the low-frequency measurements are made
with a spectrum analyzer (Stanford Research Systems SR785).

\begin{figure}[t]
\includegraphics[angle=0, width=0.9\columnwidth]{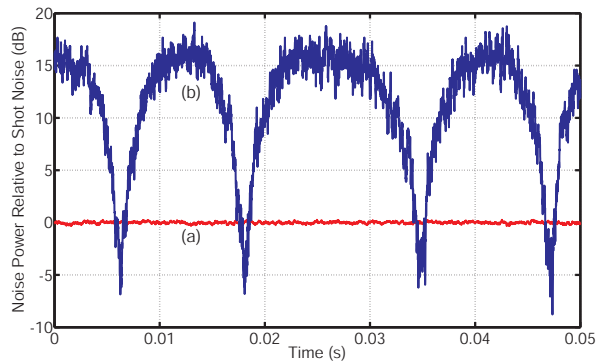}
\caption{(Color online) The fixed-frequency spectra of (a) shot
noise, and (b) squeezed/anti-squeezed noise power when the squeeze
angle was scanned as a function of time. The measurements were done
at 900 kHz with zero frequency span. The resolution bandwidth (RBW)
is 100 kHz and the video bandwidth (VBW) is 3 kHz. The squeeze angle
was scanned using PZT3 with a ramp function at 10 Hz. The electronic
noise was 9.2 dB below the shot noise level, and was subtracted from
the squeezing/anti-squeezing and shot noise spectra. }
\label{scannedspectra}
\end{figure}

\begin{figure}[t]
\includegraphics[angle=0, width=0.9\columnwidth]{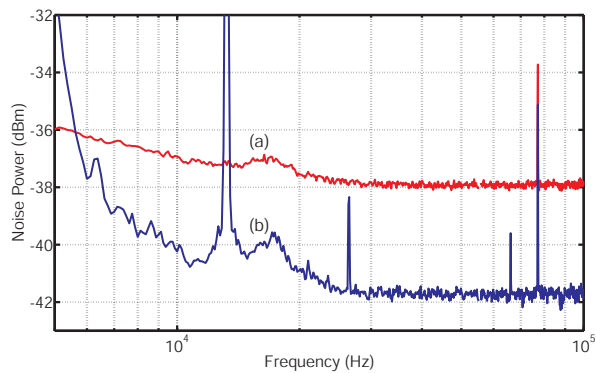}
\caption{(Color online) The broadband spectra of (a) shot noise and
(b) squeezed shot noise. The squeeze angle was noise-locked without
any coherent light. The electronic noise was 5.1 dB below the shot
noise level at low frequencies, and was subtracted from the
squeezing and shot noise spectra. The spike at 13.2 kHz and its
higher harmonic are due to the modulation of PZT3 for the
noise-locking technique.} \label{lockedspectra}
\end{figure}

We measured the fixed-frequency, zero-span spectra of the shot noise
and squeezed/anti-squeezed shot noise, shown in
Fig.~\ref{scannedspectra}, as well as the electronic noise, when
scanning the squeeze angle. The periodic oscillation of the
squeeze angle can be seen in the figure. We also measured the
spectra of the squeezed and anti-squeezed shot noise versus
frequency when the squeeze angle was noise-locked. The result is
shown in Fig. \ref{lockedspectra}. Broadband squeezing of $3.8\pm0.1$ dB at
frequencies above 30 kHz, and a cutoff frequency for squeezing at 6
kHz, were observed. The noise increase at low frequencies is due to
scattering of the LO light from the homodyne detector. To observe
squeezing at lower frequencies requires careful shielding of the LO
or ambient light induced scattered photons from coupling to the OPO
cavity. This was demonstrated very well in
Refs. \cite{mckenzie2004prl} and \cite{vahlbruch2006prl}, and is not
part of the initial goals of the present experimental demonstration.

In conclusion, we have demonstrated the generation of a stable cw
low-frequency squeezed vacuum field with a squeezing level of
$3.8\pm0.1$ dB at 1064 nm, suitable for laser interferometers for GW
detection, using PPKTP in a sub-threshold OPO. The large squeezing
levels and excellent stability of PPKTP make it an excellent
material for use in squeezing-enhanced GW detectors. The attained
squeezing level at low frequencies is limited by the availability of
higher pump power and lower intra-cavity losses. Since PPKTP is less
susceptible to photo-refractive damage and thermal lensing, a higher
squeezing level can be achieved with higher pump power and an OPO
output coupling mirror with a lower reflectivity. For the moderate
pump power level and vacuum seeding used, we did not observe
gray-tracking, a known problem for KTP.

We would like to thank our colleagues at the LIGO Laboratory,
especially C. Wipf for design and fabrication of the balanced
photodetector and K. McKenzie at Australian National University for
valuable comments. We also thank A. Furusawa at University of Tokyo
for encouraging use of PPKTP. We gratefully acknowledge support from
National Science Foundation Grant Nos. PHY-0107417 and PHY-0457264.


\end{document}